# Observation of orbital-angular-momentum-driven temperature modulation via the spin Peltier effect

Sang J. Park[1,*], Tsuneyoshi Kan[2], Takamasa Hirai[1,2] and Ken-ichi Uchida[1,2,*]

[1] National Institute for Materials Science, Tsukuba 305-0047, Japan

[2] Department of Advanced Materials Science, Graduate School of Frontier Sciences, The University of Tokyo, Kashiwa 277-8561, Japan

*Corresponding authors: PARK.SangJun@nims.go.jp (S.J.P.); UCHIDA.Kenichi@nims.go.jp (K.U.)

## Abstract

Angular-momentum transport provides a pathway for controlling energy flow in solids beyond conventional charge-based mechanisms. While spin currents are known to mediate spin-caloritronic phenomena such as the spin Peltier effect (SPE), the role of orbital angular momentum in heat transport remains largely unexplored. Here we demonstrate orbital-angular-momentum-driven temperature modulation via the SPE in yttrium iron garnet/Pt/$CuO_x$ heterostructures. Using wedge-shaped $CuO_x$ layers combined with spatially resolved active thermal measurement techniques, we map the continuous thickness dependence and quantitatively disentangle the spin- and orbital-current-mediated contributions within a single device. The orbital-mediated component exhibits a pronounced maximum at an intermediate thickness, revealing a characteristic length scale for interfacial orbital-angular-momentum generation and propagation at the Cu/$CuO_x$ interface. These results provide direct experimental evidence that charge-current-driven orbital angular momentum can drive SPE-induced temperature modulation, establishing interfacial orbital processes as an additional channel for heat transport and providing a pathway toward spin-orbit caloritronics.

## INTRODUCTION

Controlling the flow of energy in solids is central to next-generation information and energy technologies (*1*, *2*). Beyond conventional charge and phonon transport, angular momentum has emerged as a fundamental degree of freedom governing nonequilibrium dynamics in condensed matter systems. In spintronics, the generation and manipulation of spin currents, or fluxes of spin angular momentum, enable electrical control of magnetism through spin-orbit coupling (SOC) (*3*, *4*), thereby establishing angular momentum as an active channel for controlling material functionality.

When coupled with heat, angular-momentum dynamics give rise to spin-caloritronic phenomena (*5*–*7*). Representative examples include the spin Seebeck effect (SSE) (*8*, *9*) and the spin Peltier effect (SPE) (*10*, *11*), which are related through Onsager reciprocity and enable interconversion between heat and spin currents at magnetic interfaces. These discoveries have established spin angular momentum as a key mediator of temperature modulation in magnetic heterostructures. To date, most experimental realizations rely on heavy metals (HMs) with strong SOC such as Pt (*12*, *13*), reflecting a spin-centric framework in which spin currents are assumed to be the primary carriers of thermo-spin conversion.

Recent advances in orbitronics challenge this spin-centric paradigm by identifying orbital angular momentum (OAM) as a distinct transport channel in solids (*14*–*23*). The orbital Hall effect (OHE) (*16*–*18*) and orbital Rashba–Edelstein effect (OREE) (*19*–*21*) demonstrate that charge currents can generate orbital transport in bulk materials and orbital accumulation at inversion-symmetry-broken interfaces, respectively. Theoretically, such orbital responses originate from momentum-space orbital textures and are predicted to be widespread in solids, often exceeding their spin counterparts in magnitude (*14*, *24*). In materials with finite SOC, these orbital responses can be partially converted into spin currents through bulk or interfacial spin–orbit interactions (*16*, *18*), providing a pathway for orbital-to-spin angular-momentum conversion.

Although orbital transport has been studied mainly in the context of magnetic switching and torque generation (*15*, *25*, *26*), recent studies have begun to explore its role in thermally driven transport. Theoretical studies have predicted thermally driven orbital transport phenomena in magnetic systems (*27*, *28*). Experimentally, the orbital Seebeck effect, an orbital counterpart of the SSE, has been demonstrated in ferrimagnetic insulators through thermal magnons (*29*) and in nonmagnetic insulators through chiral phonons (*30*). However, whether charge-current-driven OAM can induce temperature modulation via the SPE remains unexplored.

Here we report the observation of OAM-driven temperature modulation via the SPE in ferromagnetic or ferrimagnetic insulator (FMI)/HM/orbital-active material (OM) heterostructures (**Fig. 1**). Using yttrium iron garnet (YIG; FMI)/Pt (HM)/naturally oxidized Cu ($CuO_x$; OM) trilayers, we observe a pronounced temperature modulation associated with the introduction of the $CuO_x$ layer. By continuously tuning the $CuO_x$ thickness ($t_{CuOx}$=0–16 nm) in wedge-shaped structures and spatially resolving the resulting temperature modulation using lock-in thermography (LIT), we identify a well-defined thickness at which the temperature modulation is maximized. At this optimal thickness, the orbital-current-mediated contribution reaches up to approximately three times the conventional spin-Hall-driven SPE response in Pt, demonstrating efficient charge-to-heat conversion mediated by OAM. The decrease in modulation efficiency with increasing $t_{CuOx}$ indicates that interfacial coupling involving the $CuO_x$ layer plays a key role in the observed behavior, supporting its interfacial origin associated with OREE. These findings provide direct evidence that interfacial OAM generation and transport contribute to SPE-induced temperature modulation, thereby extending the framework of spin-orbit caloritronics beyond conventional spin-dominated paradigms.

## RESULTS

### Experimental system

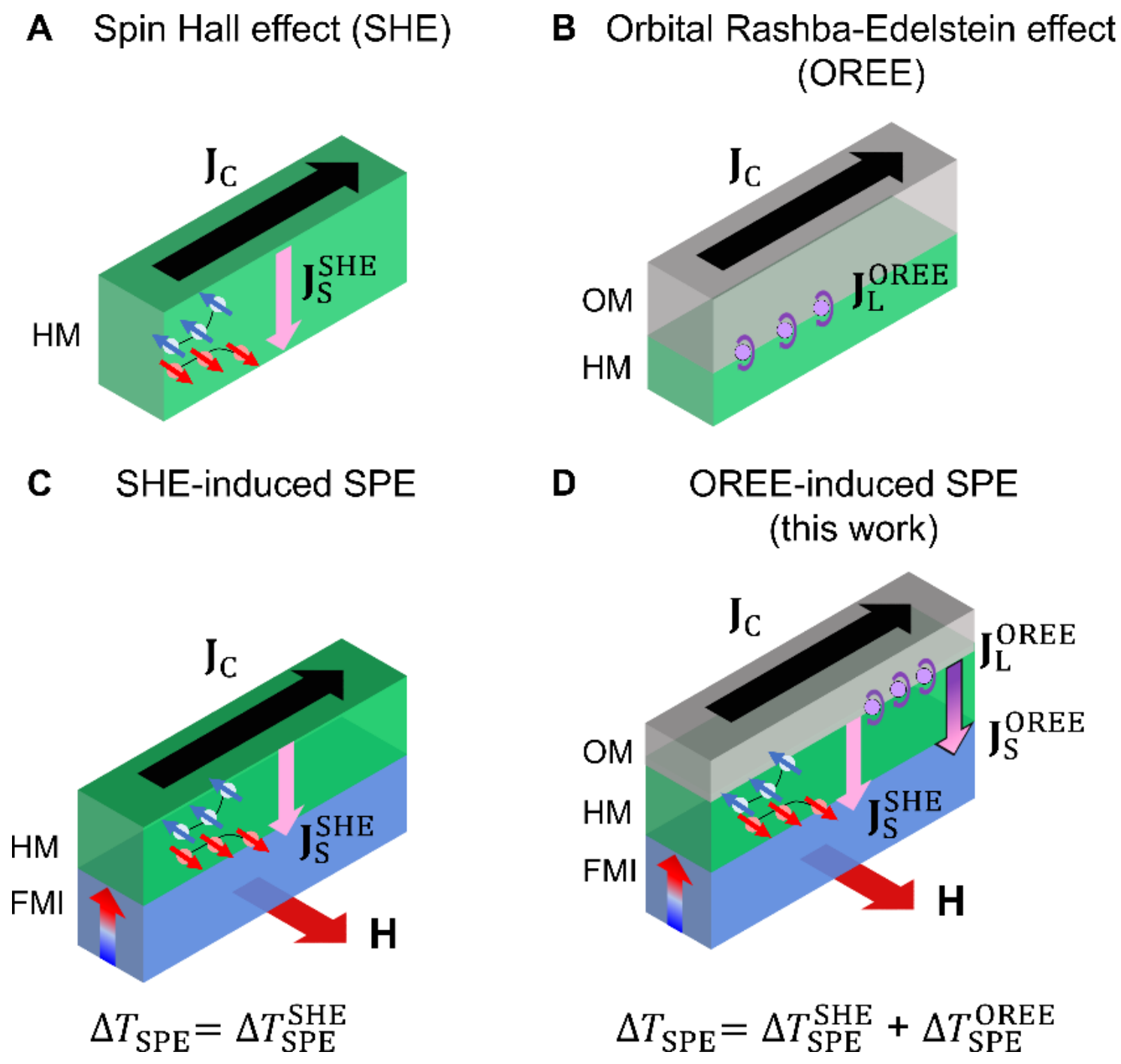

**Fig. 1. Conceptual illustration of spin- and orbital-angular-momentum-driven temperature modulation via the spin Peltier effect.** (A) Generation of a spin current $\mathbf{J}_\mathrm{S}^\mathrm{SHE}$ via the spin Hall effect (SHE) in a heavy metal (HM) with large spin–orbit coupling (SOC). (B) Generation of interfacial orbital angular momentum, followed by orbital-current propagation $\mathbf{J}_\mathrm{L}^\mathrm{OREE}$, at the HM/orbital-active material (OM) interface with broken inversion symmetry via the orbital Rashba–Edelstein effect (OREE). (C) Temperature modulation through the conventional SPE ($\Delta T_\mathrm{SPE}$) in a ferromagnetic insulator (FMI)/HM bilayer, driven solely by the spin current $\mathbf{J}_\mathrm{S}^\mathrm{SHE}$ (i.e., $\Delta T_\mathrm{SPE} = \Delta T_\mathrm{SPE}^\mathrm{SHE} \propto J_\mathrm{S}^\mathrm{SHE}$). (D) Orbital-angular-momentum-driven SPE in an FMI/HM/OM trilayer, driven by both $\mathbf{J}_\mathrm{S}^\mathrm{SHE}$ and the orbital current $\mathbf{J}_\mathrm{L}^\mathrm{OREE}$. The $\mathbf{J}_\mathrm{L}^\mathrm{OREE}$ is converted into a spin current $\mathbf{J}_\mathrm{S}^\mathrm{OREE}$ in the HM via SOC. The resulting temperature modulation is given by $\Delta T_\mathrm{SPE} = \Delta T_\mathrm{SPE}^\mathrm{SHE} + \Delta T_\mathrm{SPE}^\mathrm{OREE}$, with $\Delta T_\mathrm{SPE}^\mathrm{OREE} \propto J_\mathrm{S}^\mathrm{OREE} \propto J_\mathrm{L}^\mathrm{OREE}$.

**Figure 1** illustrates the conceptual framework of the present study. In a conventional FMI/HM bilayer, a charge current $\mathbf{J}_\mathrm{C}$ flowing in HM generates a transverse spin current $\mathbf{J}_\mathrm{S}^\mathrm{SHE}$ via the SHE (**Fig. 1A**) (*3*). Subsequently, the injected spin current exchanges angular momentum with magnons at the FMI/HM interface, producing interfacial heat absorption or release through the SPE (**Fig. 1C**) (*10*, *11*).

To examine whether orbital currents can similarly drive angular-momentum transfer responsible for temperature modulation, we introduced an orbital-active material (OM) layer on top of HM, forming FMI/HM/OM trilayers (**Fig. 1D**). At interfaces involving this layer, inversion symmetry breaking enables the generation of nonequilibrium OAM via the OREE (**Fig. 1B**), an orbital analogue of the Rashba-Edelstein effect (*31*, *32*). The generated OAM can propagate toward HM as an orbital current $\mathbf{J}_{\mathrm{L}}^{\mathrm{OREE}}$, where it is subsequently converted into a spin current $\mathbf{J}_{\mathrm{S}}^{\mathrm{OREE}}$ through SOC, thereby providing an additional contribution to interfacial angular-momentum injection and the resulting SPE alongside the conventional $\mathbf{J}_{\mathrm{S}}^{\mathrm{SHE}}$.

To experimentally implement this concept, we employed YIG, Pt, and $CuO_x$ as the FMI, HM, and OM layers, respectively. To systematically examine the orbital contribution, we fabricated a wedge-shaped $CuO_x$ layer with continuously varying thickness ($t_{CuOx}$) on a uniform YIG/Pt bilayer (**Fig. 2**, **MATERIALS AND METHODS**), covering a $t_{CuOx}$ range of 0–16 nm. This geometry enables $t_{CuOx}$-dependent evaluation without batch-to-batch variations in thin-film growth and measurement conditions. Because the $CuO_x$ layer is formed through natural oxidation of deposited Cu, the layer consists of coexisting metallic and oxidized Cu regions with a $t_{CuOx}$-dependent distribution, which is expected to promote OAM generation at the Cu/$CuO_x$ interface (*15*, *20*, *33–35*). All uncapped samples were exposed to ambient air for 2 days, after which electrical contacts were formed using silver epoxy, which was cured at 80℃ for 30 min before the LIT measurements. A time-dependent resistance measurement using the same processing sequence showed a rapid initial resistance increase during ambient exposure and a further change following the 80℃ curing treatment (**fig. S1**). Throughout this manuscript, we refer to this mixed Cu/$CuO_x$ structure as an effective OM layer and denote it as $CuO_x$. Accordingly, $t_{CuOx}$ represents the total thickness of the layers containing both metallic and oxidized Cu regions.

We used a 72-μm-thick YIG grown by liquid-phase epitaxy on a gadolinium gallium garnet (GGG) substrate (**MATERIALS AND METHODS**). This thickness is much larger than both the magnon energy relaxation length (~250 nm) (*36*, *37*) and the magnon diffusion length (~10 μm) (*38*) at room temperature, enabling bulk-like magnetic behavior without boundary or diffusion limitations. Accordingly, the GGG substrate is not explicitly discussed throughout the manuscript. The Pt thickness was fixed at 2.5 nm, which provides robust SOC-mediated angular-momentum conversion within the layer (*18*).

**Thickness-dependent SPE-induced temperature modulation**

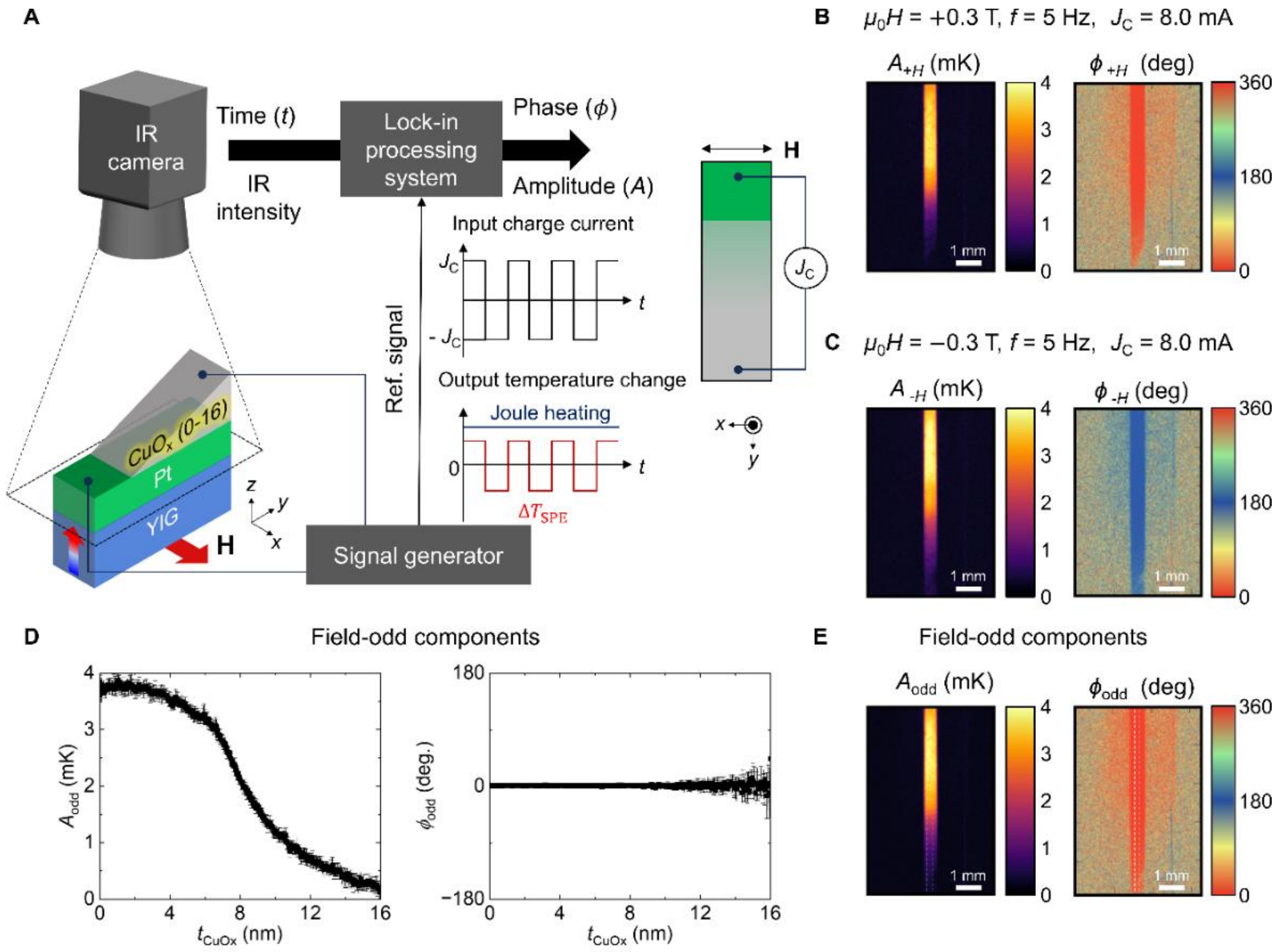

**Fig. 2. Observation of angular-momentum-driven temperature modulation via the SPE using lock-in thermography (LIT).** (A) Observation of SPE-induced temperature modulation using LIT in a $CuO_x$ wedge layer fabricated on an yttrium iron garnet (YIG)/Pt bilayer. The $CuO_x$ thickness ($t_{CuOx}$) was continuously varied from 0 to 16 nm in a single sample and the signal was spatially resolved using an infrared camera with high spatial resolution and temperature sensitivity. A square-wave-modulated current with zero offset was applied to isolate the SPE-induced temperature modulation (see **MATERIALS AND METHODS**). An external magnetic field **H** was applied along the *x*-axis to satisfy the symmetry condition for detecting the SPE-induced temperature modulation along the *z*-direction using LIT. (B and C) Measured lock-in amplitude and phase under magnetic fields of (B) $\mu_0H$ = +0.3 T and (C) $\mu_0H$ = -0.3 T, respectively, at a frequency of $f$ = 5 Hz and a current amplitude of $J_C$ = 8.0 mA. The corresponding quantities are denoted by $A_{+H}$ and $\phi_{+H}$ in (B), and by $A_{-H}$ and $\phi_{-H}$ in (C). The LIT images were acquired using a 1× infrared lens, corresponding to a pixel size of 15 × 15 $\mu m^2$. (D) Line profiles of the field-odd lock-in amplitude $A_{odd}$ and phase $\phi_{odd}$ as functions of $t_{CuOx}$. The $A_{odd}$ and $\phi_{odd}$ values represent spatial averages along the *x*-direction over the region between the dashed lines in (E). The error bars indicate the standard deviation. (E) Spatial maps of the field-odd lock-in amplitude $A_{odd}$ and phase $\phi_{odd}$, obtained from complex antisymmetrization of the LIT signals measured in (B) and (C), as defined in the **MATERIALS AND METHODS**.

The SPE-induced temperature modulation was detected using LIT under a square-wave-modulated AC charge current with zero DC offset and a steady magnetic field $H$ (**Fig. 2A**) (*39*, *40*). The infrared camera measured the amplitude $A$ and phase $\phi$ of the temperature oscillation relative to the applied current $J_C$. To ensure sufficiently high and spatially uniform infrared emissivity, the surfaces of all samples were coated with black ink having an emissivity greater than 0.94 (**fig. S2**). The patterned line was oriented along the $t_{CuOx}$-gradient direction ($y$), enabling continuous $t_{CuOx}$-dependent evaluation of the SPE-induced temperature modulation under a constant current density $j_C$. The measurements were performed at a modulation frequency of $f = 5$ Hz and opposite saturating magnetic fields of $\mu_0 H = \pm 0.3$ T, where $\mu_0$ is the vacuum permeability.

**Figures 2B** and **2C** show the $A$ and $\phi$ images measured at $\mu_0 H = +0.3$ T and -0.3 T, respectively. Upon reversal of the magnetic field, the lock-in amplitude remained similar, whereas the phase in the current-carrying region changed by approximately 180°, indicating reversal of the complex temperature-modulation signal with the YIG magnetization. To extract the magnetization-dependent, field-odd component associated with spin and/or OAM transfer, the complex LIT signals were antisymmetrized according to $Z_{\mathrm{odd}} = [Z(+H) - Z(-H)]/2$ (see **MATERIALS AND METHODS**). Because Joule heating ($\propto J_C^2$) yields a constant background without a first-harmonic response and the conventional Peltier effect ($\propto J_C$) is field-independent, this procedure isolates the SPE-induced temperature modulation that is linear in $J_C$ and odd with respect to magnetization (*11*, *41*).

**Figure 2E** shows the resulting spatial maps of the field-odd lock-in amplitude $A_{\mathrm{odd}}$ and phase $\phi_{\mathrm{odd}}$, obtained from the raw signals shown in **Figs. 2B** and 2**C**. The LIT images had a pixel size of 15×15 μm$^2$, corresponding to an effective $t_{CuOx}$ sampling interval of 0.04154 nm per pixel along the wedge gradient. The corresponding $t_{CuOx}$-dependent profiles shown in Fig. 2D were obtained by averaging the spatial maps along the constant-$t_{CuOx}$ direction ($x$ direction). The field-even components were negligibly small, as shown in **fig. S3**.

**Figure 2D** shows that $A_{\mathrm{odd}}$ remains nearly constant up to $t_{CuOx} \sim 4$ nm, followed by a gradual and monotonic decrease with increasing $t_{CuOx}$. The corresponding $\phi_{\mathrm{odd}}$ remains stable within the finite-signal regime and becomes noisy only when $A_{odd}$ approaches the sub-mK regime at larger $t_{CuOx}$. Importantly, no systematic phase shift or sign reversal is observed over the entire finite-signal regime. The constant $\phi_{odd}$ suggests that the SPE contributions associated with Pt and $CuO_x$ share the same sign and symmetry with respect to magnetization and charge-current direction, as no sign reversal is observed over the measured thickness range, consistent with previous observations in TmIG/Pt/$CuO_x$ systems (*15*). The systematic suppression of $A_{odd}$ at larger $t_{CuOx}$ suggests a reduction of the effective angular-momentum injection at the YIG/Pt interface, potentially associated with (1) current redistribution within the multilayer and/or (2) modification of interfaces involving the $CuO_x$ layer responsible for OAM generation.

To further characterize the magnetic-field and current dependences of the temperature modulation, we performed additional LIT measurements using an isolated YIG/Pt(2.5 nm)/$CuO_x$(5.6 nm) device. $A$ and $\phi$ were measured at magnetic-field magnitudes ranging from 0.08 to 0.30 T under both field polarities. Within the saturated-field regime, the amplitude remained nearly constant within the experimental uncertainty, whereas the phase reversed with the magnetic-field polarity (**fig. S4**). These results show that the temperature-modulation signal follows the YIG magnetization orientation. In addition, current-dependent measurements showed that $A_{odd}$ increased linearly with $J_C$ ($R^2 > 0.9999$) (**fig. S5**). The field-odd symmetry and linear current dependence support the assignment of the observed field-odd temperature modulation to the SPE and distinguish it from the field-independent conventional Peltier effect and $J_C^2$-dependent Joule heating contribution. Because the observed temperature modulation contains contributions from both the conventional SHE in Pt and the orbital-mediated component associated with $CuO_x$, quantitative separation of these contributions is required to clarify the role of OAM in SPE-induced temperature modulation.

## Quantitative separation of spin- and orbital-current contributions

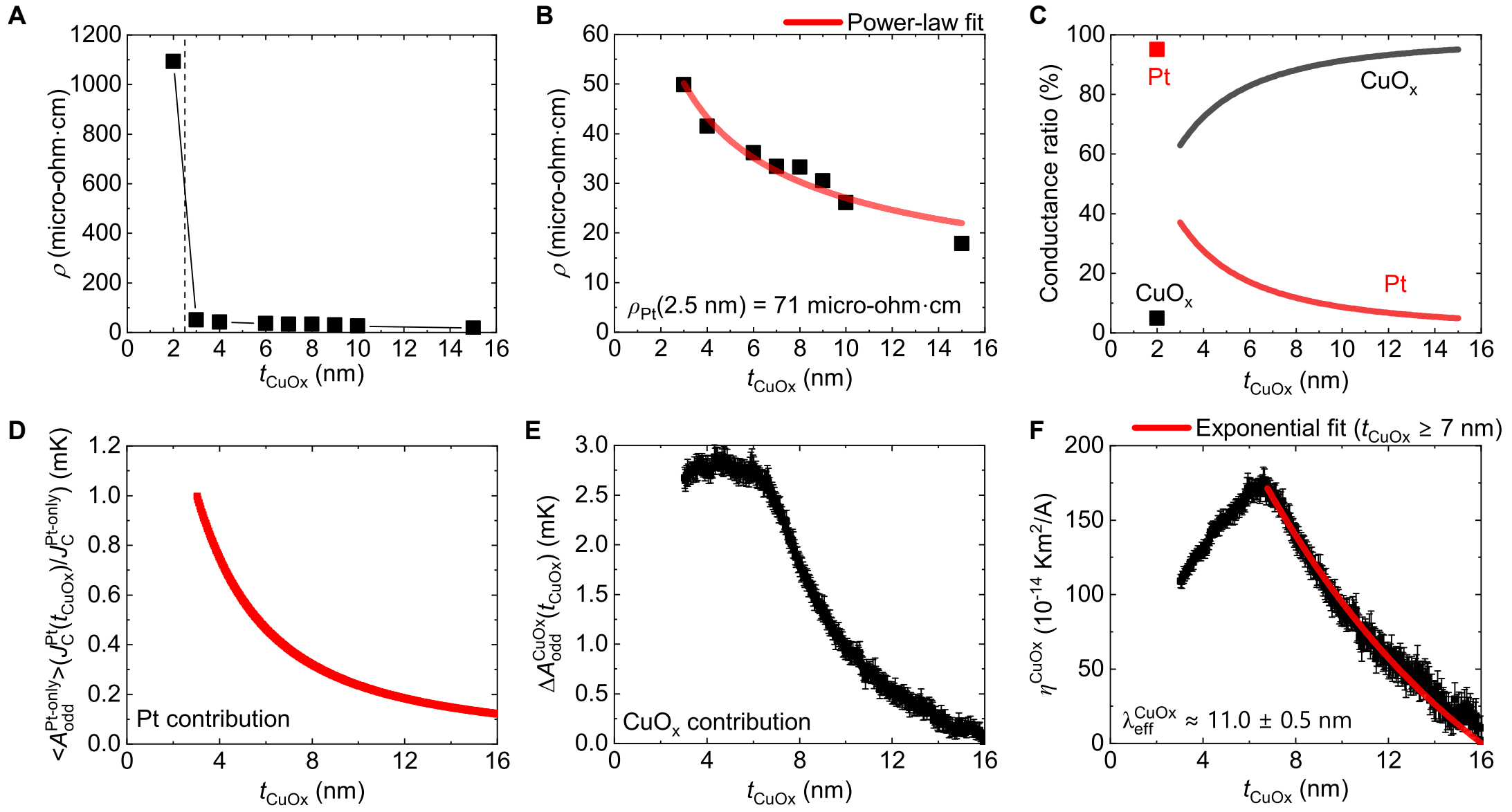

**Fig. 3. Quantitative separation of SPE contributions from Pt and $CuO_x$ based on bilayer electrical conductance.** (A) Electrical resistivity $\rho$ of uniform $CuO_x$ films as a function of $CuO_x$ thickness ($t_{CuOx}$), measured using the standard four-probe method (inset). The dashed line indicates the percolation limit. (B) Selected view of $\rho$ for 3–15 nm thick films. The data are fitted using a phenomenological power-law function, $\rho(t_{CuOx}) = \rho_\infty + C t_{CuOx}^{-n}$, where $\rho_\infty$ is the bulk resistivity limit of Cu, and $C$ and $n$ are fitting parameters. The $\rho$ of 2.5-nm-thick Pt was measured to be 71 micro-ohm·cm. The error bars for (A) and (B) are smaller than the symbols. (C) Electrical conductance ratio of $CuO_x(t_{CuOx})$ and Pt(2.5 nm) as a function of $t_{CuOx}$, used to determine the current distribution within the multilayer. (D and E) Separated SPE-induced temperature modulation from (D) Pt and (E) orbital contributions associated with $CuO_x$ to the measured $A_{odd}$ signal as a function of $t_{CuOx}$ (see **Equation 2** for details). (F) Charge-to-heat conversion efficiency $\eta^{CuOx}$ associated with the $CuO_x$ orbital contribution, defined as SPE-induced temperature modulation per the current density flowing through the $CuO_x$ layer (see **Equation 3**). The decay in the $t_{CuOx} \geq 7$ nm region is fitted using an exponential function $\exp(-\frac{t_{CuOx}}{\lambda_{eff}^{CuOx}})$. The values in (F) are replotted in **Fig. 4** and **fig. S5C** for comparison.

To clarify the orbital contribution, we quantitatively separated the individual components associated with SHE and orbital-current mechanisms based on the thickness-dependent current distribution within the multilayer. We first evaluated the effective electrical resistivity $\rho$ of the $CuO_x$ films as a function of $t_{CuOx}$ using a standard four-probe method (**MATERIALS AND METHODS**, **Fig. 3A inset**). As mentioned above, $t_{CuOx}$ represents the effective thickness of the $Cu/CuO_x$ layers. Electrical resistivity exhibits strong thickness dependence in the ultrathin regime; $\rho$ increases when $t_{\mathrm{CuOx}}$ decreases because of enhanced interface scattering and reduced metallic conduction thickness. At $t_{\mathrm{CuOx}} = 2$ nm, $\rho$ increases sharply by approximately one order of magnitude to ~$1.1\times10^3$ micro-ohm·cm, indicating the percolation limit and suggesting an effective oxidation layer thickness of ~ 2–3 nm.

To estimate the current distribution in the multilayers continuously as a function of $t_{\mathrm{CuOx}}$, we described the resistivity for $t_{\mathrm{CuOx}}$ in the range 3–15 nm by a phenomenological power-law form,

$$\rho(t_{\mathrm{CuOx}}) = \rho_\infty + C t_{\mathrm{CuOx}}^{-n}, \tag{1}$$

where $\rho_\infty$ is the bulk resistivity limit of Cu (1.68 micro-ohm·cm), and $C$ and $n$ are fitting parameters (**Fig. 3B**). This fitting allows interpolation of the resistivity values for continuous current-distribution analysis in the wedged sample. The fitting parameters are $C = 88.2\pm7.9$ and $n = 0.513\pm0.051$, valid for $t_{\mathrm{CuOx}} = 3$–15 nm. The resistivity of 2.5-nm-thick Pt ($\rho_{\mathrm{Pt}}$) was measured independently as 71 micro-ohm·cm.

Using the extracted resistivity values and layer thickness, we determined the sheet conductance of Pt ($G_{\mathrm{Pt}}$) and $CuO_x$ ($G_{\mathrm{CuOx}}$) layers and calculated the thickness-dependent conductance ratio. Assuming parallel current flow within the metallic layers, the fractions of the applied current flowing through Pt and $CuO_x$ were determined from their relative conductances as $G_{\mathrm{Pt}}/G_{\mathrm{tot}}$ and $G_{\mathrm{CuOx}}/G_{\mathrm{tot}}$, where $G_{\mathrm{tot}} = G_{\mathrm{Pt}} + G_{\mathrm{CuOx}}$. **Figure 3C** shows the conductance ratio, which represents the current distribution within the valid fitting range of $t_{\mathrm{CuOx}} = 3$–15 nm. At $t_{\mathrm{CuOx}} = 3$ nm, approximately 63% of the applied current flows through the $CuO_x$ layer, and this fraction increases with increasing $t_{\mathrm{CuOx}}$. It is noted that the current flowing through a 2-nm-thick $CuO_x$ layer is merely 5% of the total applied current, suggesting that the SPE signal observed below 2 nm is primarily governed by conventional charge-to-spin conversion processes rather than by orbital transport through the ultrathin $CuO_x$ layer.

Based on this conductance model, we separated the SPE-induced temperature modulation attributed to the $CuO_x$ layer by subtracting the Pt-only contribution expected from the conventional SHE. As experimentally observed in **Fig. 2**, $\phi_{odd}$ remains constant; therefore, the complex lock-in signals add constructively without additional phase lag, allowing direct subtraction of the amplitudes. Accordingly, the $CuO_x$-associated component is obtained as

$$\Delta A_{\mathrm{odd}}^{\mathrm{CuOx}}(t_{\mathrm{CuOx}}) = \langle A_{\mathrm{odd}}^{\mathrm{Pt/CuOx}}(t_{\mathrm{CuOx}})\rangle - \langle A_{\mathrm{odd}}^{\mathrm{Pt-only}}\rangle\left(\frac{J_{\mathrm{C}}^{\mathrm{Pt}}(t_{\mathrm{CuOx}})}{J_{\mathrm{C}}^{\mathrm{Pt-only}}}\right), \quad (2)$$

where $\langle A_{\mathrm{odd}}^{\mathrm{Pt/CuOx}}(t_{\mathrm{CuOx}})\rangle$ and $\langle A_{\mathrm{odd}}^{\mathrm{Pt-only}}\rangle$ represent the spatial averages in the constant $t_{\mathrm{CuOx}}$ region (along $x$ direction) and the YIG/Pt reference region, respectively. The Pt contribution (second term on the right-hand side of Equation (2)) is shown in **Fig. 3D**, while the $CuO_x$-associated component (left-hand side) is presented in **Fig. 3E**. $J_{\mathrm{C}}^{\mathrm{Pt}}(t_{\mathrm{CuOx}})$ is the thickness-dependent current flowing through the Pt layer, estimated from the conductance model as $J_{\mathrm{C}}^{\mathrm{Pt}}(t_{\mathrm{CuOx}}) = (G_{\mathrm{Pt}}/G_{\mathrm{tot}})J_{\mathrm{C}}$ (**Fig. 3C**). The resulting $CuO_x$-associated SPE contribution was then normalized by the corresponding current density $j_{\mathrm{C}}^{\mathrm{CuOx}}(t_{\mathrm{CuOx}})$ flowing through the $CuO_x$ layer as

$$\eta^{\mathrm{CuOx}} = \Delta A_{\mathrm{odd}}^{\mathrm{CuOx}}(t_{\mathrm{CuOx}})/j_{\mathrm{C}}^{\mathrm{CuOx}}(t_{\mathrm{CuOx}}), \quad (3)$$

thereby defining a temperature-modulation efficiency associated with the orbital contribution.

Notably, the normalized amplitude exhibits a peak at $t_{\mathrm{CuOx}}$ ~ 6–7 nm (**Fig. 3F**), indicating an optimal thickness scale for the orbital-mediated response. If the signal originated from bulk orbital transport (e.g., the OHE), it would be expected to increase with thickness and eventually saturate over a scale determined by the propagation length (*17*, *42*). The pronounced peak instead suggests an interfacial origin of OAM generation, consistent with an OREE-driven SPE heat-modulation mechanism. Considering an estimated oxide-layer thickness of approximately 2 to 3 nm (**Fig. 3A**), OAM generation is likely most effective when oxidized and metallic Cu regions coexist, a configuration reported to favor the OREE (*15*, *29*, *34*, *35*, *43–46*). First-principles calculations have shown that hybridization between O 2*p* and Cu 3*d* orbitals induces a pronounced orbital Rashba effect in surface-oxidized Cu (*35*). More recently, combined XPS, TEM, EELS, and first-principles analyses identified orbital hybridization at the $Cu_4O_3$/Cu interface as a key factor in enhancing orbital-current generation in naturally oxidized $CuO_x$ heterostructures (*46*). These findings support our interpretation that the coexistence of oxidized and metallic Cu regions promotes interfacial OAM generation.

At this optimal thickness, the peak value ($177 \times 10^{-14}$ $Km^2/A$) exceeds the conventional spin-driven SPE value measured in the same device ($58 \times 10^{-14}$ $Km^2/A$) and the representative value reported for YIG/Pt ($47 \times 10^{-14}$ $Km^2/A$) (*41*), highlighting efficient charge-to-heat conversion mediated by the OAM. This wedge-based approach minimizes uncertainties arising from batch-to-batch variations and interfacial differences, enabling direct evaluation of the intrinsic thickness dependence. At larger $t_{CuOx}$, the signal monotonically decreases and becomes comparable to the uncertainties. The finite orbital contribution observed up to $t_{CuOx} > 12$ nm, where a direct Pt/$CuO_x$ interface is effectively suppressed, further suggests that the signal is associated with the Cu/CuOx interface. To further quantify this behavior, we fitted the decreasing data in the range $t_{CuOx}$ = 7–15 nm using an exponential function $\exp(-\frac{t_{CuOx}}{\lambda_{eff}^{CuOx}})$, where $\lambda_{eff}^{CuOx}$ represents an effective characteristic length. The decay is well described by the exponential function within the experimental uncertainty ($R^2 > 0.996$), yielding $\lambda_{eff}^{CuOx} = 11.0 \pm 0.5$ nm. This exponential decay indicates a diffusion-like attenuation of the interfacial orbital contribution. The extracted length scale $\lambda_{eff}^{CuOx} = 11.0 \pm 0.5$ nm is comparable to the orbital transport length scale reported in previous studies on metallic Cu, where a characteristic length of ~ 9 nm was obtained from time-resolved measurements (*47*). The observed behavior strongly supports OAM generation at the Cu/$CuO_x$ interface, while a direct Pt/$CuO_x$ contribution is unlikely, with the signal attenuating through the metallic Cu layer between Pt and $CuO_x$. As an independent check, a finite $\eta^{CuOx}$ was also obtained for a fixed-thickness isolated-line device, indicating that the orbital contribution persists without the continuous wedge geometry (**fig. S5C**).

## Origin of the orbital-current-induced temperature modulation

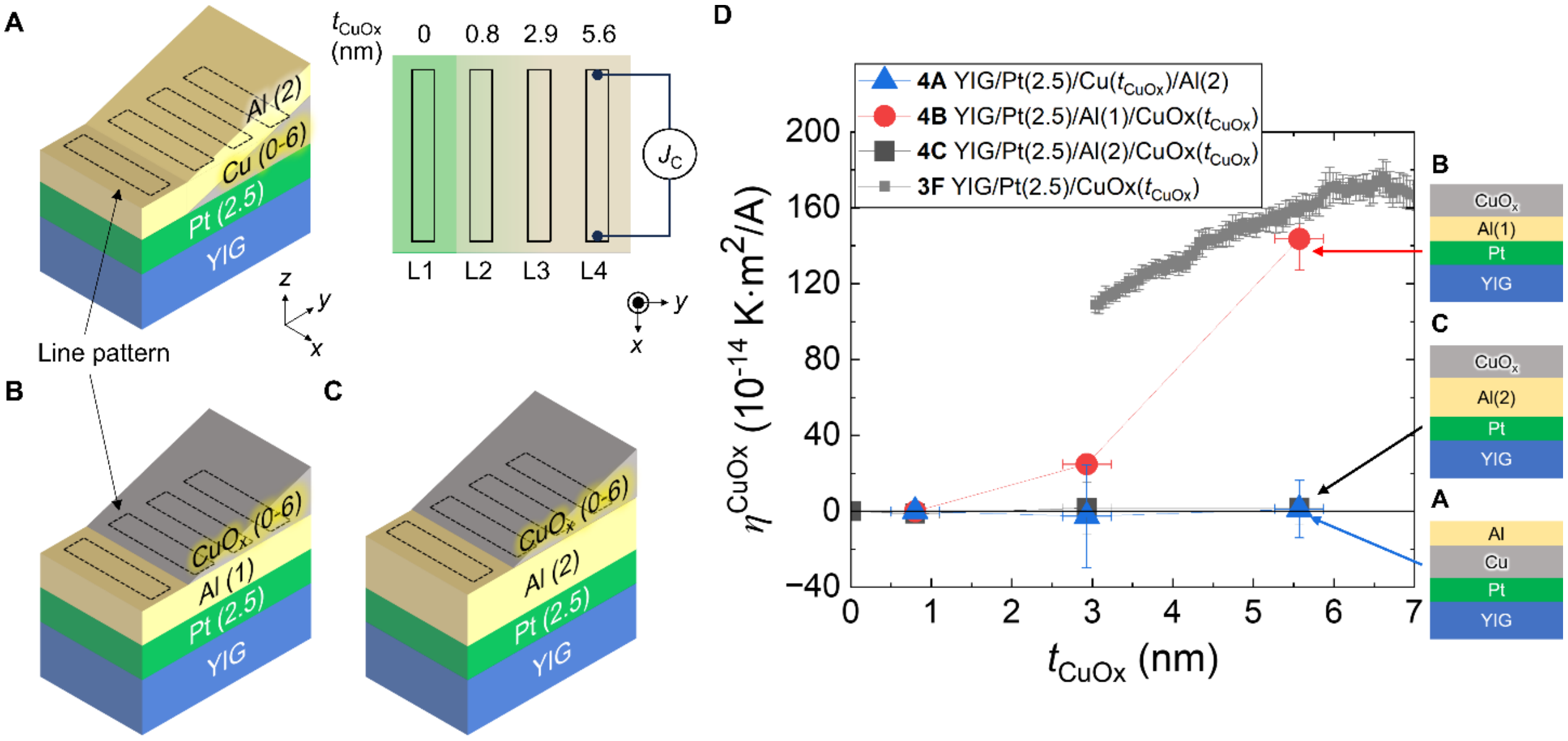

**Fig. 4. Separated orbital-angular-momentum-induced temperature modulation efficiency in samples patterned along the *x* direction, perpendicular to $CuO_x$ thickness gradient.** (A)–(C) Schematic illustrations of sample structures with (A) YIG/Pt(2.5 nm)/Cu($t_{CuOx}$) capped with a 2-nm-thick Al layer to suppress surface oxidation, (B) YIG/Pt(2.5 nm)/Al(1 nm)/$CuO_x$($t_{CuOx}$) and (C) YIG/Pt(2.5 nm)/Al(2 nm)/$CuO_x$($t_{CuOx}$), where a thin Al layer is inserted between the Pt and $CuO_x$ layers. (D) Orbital temperature modulation efficiency $\eta^{CuOx}$ for the samples (A)–(C) and the main data in **Fig. 3F** for reference. The corresponding LIT images and analysis regions are provided in **fig. S6**.

To further investigate the origin of OAM generation, we fabricated additional samples using the same procedures (see **MATERIALS AND METHODS**) with Al-capped and Al-inserted structures, as shown in **Fig. 4A–C**. The Al-capped sample (YIG/Pt(2.5)/Cu($t_{CuOx}$)/Al(2), where all thicknesses are in nm) was fabricated to suppress oxidation of Cu and thereby inhibit the formation of $CuO_x$. The Al-inserted samples (YIG/Pt(2.5)/Al(1 or 2)/$CuO_x$($t_{CuOx}$)) were fabricated to probe the origin of the signals by suppressing the Pt/$CuO_x$ interface while preserving the Cu/$CuO_x$ interface. Ultrathin Al layers (with a thickness of ~0.8 nm) have been reported to provide appreciable interfacial orbital transparency (*48*). Therefore, if the OAM is generated at the Cu/$CuO_x$ interface via the OREE and propagates toward Pt as an orbital current through the metallic Cu layer, as suggested by the results in **Fig. 3**, this Al-inserted structure is expected to exhibit a finite orbital-driven temperature modulation. The patterned lines were oriented along the *x* direction, perpendicular to the $t_{CuOx}$-gradient direction, enabling discrete evaluation at selected thicknesses. Four patterned lines, L1–L4, were fabricated corresponding to $t_{CuOx}$ = 0, 0.8, 2.9, and 5.6 nm, respectively. The variation in $t_{CuOx}$ across the width of each patterned line was estimated to be ~0.3 nm and is represented by the horizontal error bars in **Fig. 4D**. L1 was used as a reference, and L2 ($t_{CuOx}$ = 0.8 nm) exhibited nearly identical signals. All LIT measurements were conducted sequentially by attaching electrodes to each patterned line. Using the LIT measurement and quantitative signal separation described above, we extracted the orbital contribution $\eta^{CuOx}$ (**Fig. 4D**). The detailed LIT data are provided in **fig. S6**.

Interestingly, the Al-capped sample (YIG/Pt(2.5)/Cu($t_{CuOx}$)/Al(2), **Fig. 4A**) showed negligible orbital contribution (~$2\times10^{-14}$ $Km^2/A$) for both L3 and L4, well below the experimental uncertainty (4–$6\times10^{-14}$ $Km^2/A$). In these samples, the observed signal is dominated by the conventional SHE in Pt. This result indicates that in the YIG/Pt/$CuO_x$ samples used for the experiments in **Figs. 2** and **3**, $CuO_x$ plays a critical role in OAM generation through the OREE, consistent with previous theoretical and experimental studies (*15*, *29*, *34*, *35*, *43–45*). Although bulk orbital currents generated in Cu have been reported (*49*), the disappearance of the signal upon Al capping indicates that bulk contributions are not the dominant origin of the temperature modulation observed in our system. It is worth noting that both YIG/Cu(5)/Pt(5) and YIG/$Al_2O_3$(1)/Pt(5) produce negligibly small SPE signals (*41*), further indicating that pure Cu or oxidized Al layers alone do not produce a measurable SPE response in the present configuration.

We then investigated a structure with an Al layer inserted between Pt and $CuO_x$ (**Fig. 4B,C**), designed to disrupt the Pt/$CuO_x$ interface while preserving the Cu/$CuO_x$ side of the stack. With a 1-nm-thick Al layer, which is reported to provide appreciable interfacial orbital transparency (*48*), we observed a finite orbital contribution with a maximum $\eta^{CuOx}$ of 144 $\times10^{-14}$ $Km^2/A$ at L4. However, no orbital-driven signal was observed when the Al thickness was 2 nm. The persistence of the signal with the 1-nm-thick Al insertion, despite disruption of the direct Pt/$CuO_x$ interface, supports OAM generation at the Cu/$CuO_x$ interface (*46*, *50*).

Its disappearance with the 2-nm-thick Al insertion is consistent with reduced orbital-current transmission through the thicker Al layer. Together with the exponential decay trend observed in **Fig. 3**, these results strongly support that OAM generation predominantly occurs at the Cu/$CuO_x$ interface, while contributions from a direct Pt/$CuO_x$ interface are unlikely in the present configuration.

## DISCUSSION

We demonstrate that orbital currents can contribute to temperature modulation in magnetic heterostructures via the SPE. Using YIG/Pt/$CuO_x$ trilayers with continuously tuned $CuO_x$ thickness, we identify an orbital-current-mediated contribution that becomes pronounced under optimized conditions. The thickness-dependent behavior, characterized by a distinct maximum followed by exponential decay, together with control experiments using Al-capped and Al-inserted structures, indicates that the observed temperature modulation is primarily associated with interfacial OAM generation at the Cu/$CuO_x$ interface. The suppression in the Al-capped structure and the persistence of a finite signal with 1-nm-thick Al insertion highlight the essential role of interfacial coupling, ruling out bulk orbital transport as the dominant mechanism. These results establish a direct link between charge-current-driven OAM transport and SPE-induced temperature modulation, extending the scope of spin caloritronics beyond spin-dominated frameworks. In addition, the interplay between orbital and spin currents suggests a potential driving principle for transverse thermoelectric conversion, where angular-momentum transport can be utilized for heat-to-charge energy conversion. In this context, orbital contributions may provide an additional degree of freedom to complement established spin-based mechanisms (*51*, *52*). Future observation of the orbital Peltier effect, the Onsager reciprocal counterpart of the orbital Seebeck effect, would further broaden the framework of orbital-mediated heat transport. The ability to engineer interfacial OAM generation and transport provides a design principle for heat control in solids, thereby opening a pathway toward spin-orbit caloritronics and highlighting the potential for further performance improvement in solid-state thermal management and energy conversion technologies.

## MATERIALS AND METHODS

### Sample preparation

The heterostructure samples were fabricated by depositing thin metallic layers using a magnetron sputtering system (Comet Inc., CMS-A6250X2) on (111)-oriented single-crystalline YIG grown on (111)-oriented GGG substrates by liquid-phase epitaxy. The GGG/YIG substrates were mirror-polished before deposition using a 50-nm-diameter alumina slurry (Baikowski International Corporation) and cleaned by sonication in acetone and deionized water to remove residues. Direct-current power sources were used for the Pt, Cu, and Al targets with a power of 30 W, and the deposition rates were 0.0349, 0.0339, and 0.0123 nm/s, respectively. The deposition was conducted at room temperature under vacuum conditions with a base pressure below $6.0 \times 10^{-6}$ Pa and at an Ar process gas pressure of 0.5 Pa. Uniform films of Pt and Al were deposited while rotating the substrates, whereas the wedged Cu layer was fabricated using a linearly moving shutter over a length of approximately 6 mm at a fixed substrate angle. The layers were deposited without breaking vacuum. All samples were exposed to ambient air for 2 days before further processing. In samples without a capping layer, this exposure resulted in the formation of naturally oxidized $CuO_x$ layers from the deposited Cu films, whereas the Al-capped sample was protected from oxidation. For the LIT samples, electrical contacts were formed at the edges of the patterned lines using silver epoxy, which was cured at 80℃ for 30 min. The samples used for electrical resistivity measurements were subjected to the same thermal treatment before measurement.

### Lock-in thermography measurement and field antisymmetrization

The spatial distribution of temperature modulation associated with the SPE was measured using a LIT system (DCG Systems Inc., ELITE). Before the LIT measurements, all sample surfaces were uniformly coated with black ink having an emissivity greater than 0.94. For the continuous wedge sample shown in **Fig. 2**, a 1× infrared lens with an f-number of 1.5 and a working distance of 40 mm was used, corresponding to a pixel size of $15 \times 15$ μm$^2$. The total image dimensions were 329 × 486 pixels, and the quantitative analysis was performed over a region of 17 × 486 pixels. For the measurements corresponding to Fig. 4 and shown in **fig. S6**, a 5× infrared lens with an f-number of 4.0 and a working distance of 14 mm was used, corresponding to a pixel size of 3 × 3 μm². The total image dimensions were 446 × 304 pixels, and the quantitative analysis was performed over a region of 91 × 304 pixels.

To isolate the magnetization-dependent SPE component, the signals were antisymmetrized with respect to the magnetic field (±*H*). The complex LIT signal measured under each magnetic-field direction was defined as

$$Z_{\pm H} = A_{\pm H} \exp(-i\phi_{\pm H}), \tag{4}$$

where $A_{\pm H}$ and $\phi_{\pm H}$ denote the lock-in amplitude and phase measured under positive or negative *H*, respectively. The field-odd complex LIT signal was obtained by complex antisymmetrization as follows (*11*, *39*, *40*):

$$Z_{\text{odd}} = \frac{Z_{+H} - Z_{-H}}{2}. \tag{5-1}$$

The corresponding field-odd amplitude and phase, $A_{\text{odd}}$ and $\phi_{\text{odd}}$, were determined from

$$A_{\text{odd}} = |Z_{\text{odd}}| = \left|\frac{A_{+H}\exp(-i\phi_{+H}) - A_{-H}\exp(-i\phi_{-H})}{2}\right|, \tag{5-2}$$

$$\phi_{\text{odd}} = -\arg(Z_{\text{odd}}) = -\arg\left[\frac{(A_{+H}\exp(-i\phi_{+H}) - A_{-H}\exp(-i\phi_{-H}))}{2}\right]. \tag{5-3}$$

Similarly, the field-even complex LIT signal was obtained by complex symmetrization:

$$Z_{\text{even}} = \frac{Z_{+H} + Z_{-H}}{2}. \tag{6-1}$$

The corresponding field-even amplitude and phase, $A_{\text{even}}$ and $\phi_{\text{even}}$, were determined from

$$A_{\text{even}} = |Z_{\text{even}}| = \left|\frac{A_{+H}\exp(-i\phi_{+H}) + A_{-H}\exp(-i\phi_{-H})}{2}\right|, \tag{6-2}$$

$$\phi_{\text{even}} = -\arg(Z_{\text{even}}) = -\arg\left[\frac{(A_{+H}\exp(-i\phi_{+H}) + A_{-H}\exp(-i\phi_{-H}))}{2}\right]. \tag{6-3}$$

Because the conventional Peltier effect is independent of *H* and the magneto-Peltier effect is symmetric with respect to *H*, their contributions do not appear in the field-odd amplitude and phase, $A_{\text{odd}}$ and $\phi_{\text{odd}}$.

### Electrical transport measurements

The resistivity was calculated from the measured resistance using the geometrical dimensions of the channel at room temperature. The distance between the voltage electrodes was 4.4 mm, and the channel width was 0.3 mm. The samples for electrical resistivity estimation were fabricated using the same procedure as the main samples (described above). A standard four-probe method was used with a digital multimeter (Keithley, DMM6500) and a probe system equipped with four needle probes. The measured voltage fluctuation was less than 0.1% for all samples.

For the time-dependent electrical characterization, the resistance of a YIG/Pt(2.5 nm)/Cu(6 nm) reference sample was measured in a four-probe configuration using the same digital multimeter. The measurement was initiated within 20 min after the sample was removed from the sputtering chamber and was performed at 10-min intervals. The sample was maintained under ambient conditions for 48 h and subsequently heated on a hot plate at 80℃ for 30 min to reproduce the silver-epoxy curing treatment used for the LIT samples. After the sample cooled to room temperature, resistance monitoring was resumed for a further 24 h.

**Acknowledgments**

The authors thank Hojun Lee, Jongjun M. Lee and Hyun-Woo Lee at POSTECH for valuable discussions, Yuya Sakuraba at NIMS for providing access to the sputtering system, and Mizue Isomura for technical support.

**Funding:**

This work was supported by ERATO "Magnetic Thermal Management Materials" (grant no. JPMJER2201) from JST, Japan and Grant-in-Aid for Scientific Research (S) (22H04965) from JSPS, Japan.

**Author contributions:**

S.J.P. conceived and designed the study. K.U. supervised the project. S.J.P. fabricated the films with assistance from T.K. and T.H. S.J.P. measured and analyzed the spin- and orbital-current-driven SPE temperature modulation using LIT. S.J.P. wrote the manuscript with input from all authors.

**Competing interests:**

The authors declare they have no competing interests.

**Data, code, and materials availability:**

All data needed to evaluate and reproduce the results in the paper are present in the paper and/or the Supplementary Materials. This study did not generate new materials.

# Supplementary Materials for

## Observation of orbital-angular-momentum-driven temperature modulation via the spin Peltier effect

Sang J. Park *et al.*

*Corresponding author. PARK.SangJun@nims.go.jp (S.J.P.); UCHIDA.Kenichi@nims.go.jp (K.U.)

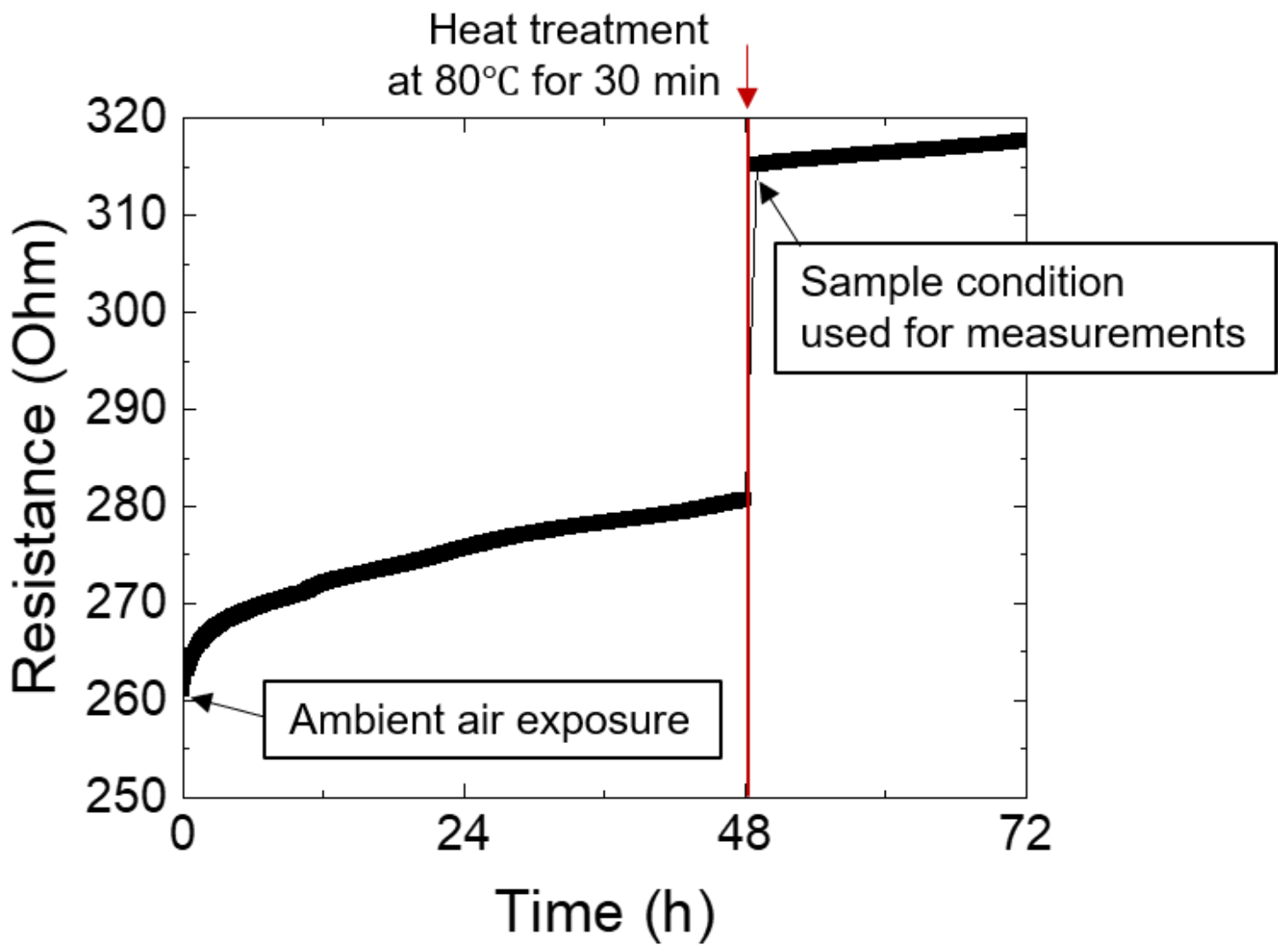

**Fig. S1. Time-dependent resistance of a YIG/Pt(2.5 nm)/Cu(6 nm) reference sample.** The resistance measurement was initiated within 20 min after the sample was removed from the sputtering chamber and was performed at 10-min intervals. The sample was maintained under ambient conditions for 48 h and subsequently heated at 80℃ for 30 min to reproduce the silver-epoxy curing treatment. After the sample cooled to room temperature, resistance monitoring was resumed for a further 24 h. The resistance increased rapidly during the initial ambient exposure, particularly during the first 5 h, and subsequently increased at a reduced rate. Following the 80°C treatment, a further resistance change occurred, and the resistance continued to increase slightly during the subsequent 24 h at room temperature.

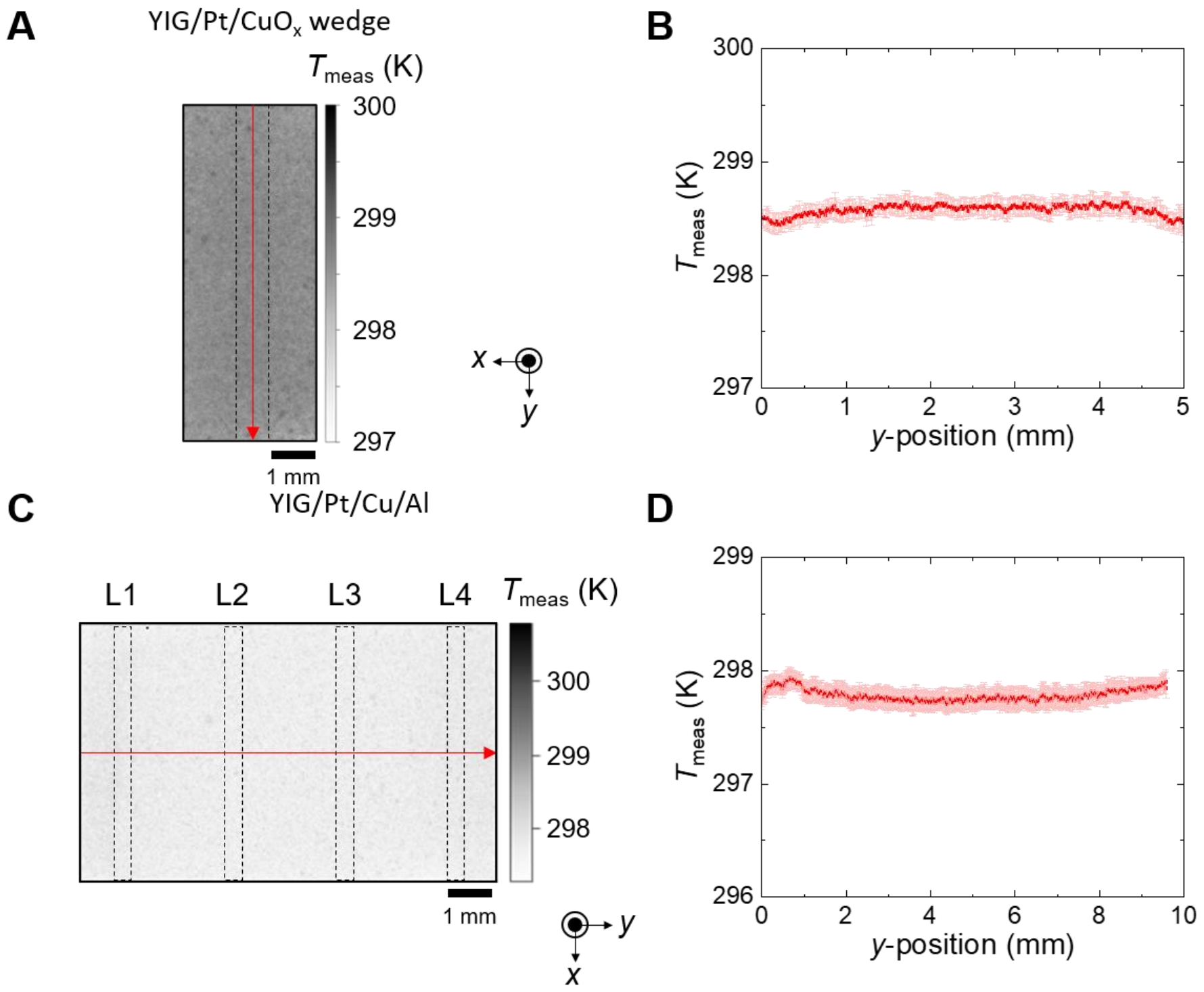

**Fig. S2. Steady-state surface-temperature distributions measured without an applied charge current.** (A) Surface-temperature map of the YIG/Pt/$CuO_x$ wedge sample, corresponding to Fig. 2 in the main text. (B) Corresponding line profile along the *y* direction indicated by the red arrow in (A). (C) Surface-temperature map of the YIG/Pt/Cu/Al sample, corresponding to Fig. 4 in the main text. (D) Corresponding line profile along the *y* direction indicated by the red arrow in (C). The white dashed boxes indicate the sample regions. No apparent temperature contrast was observed between the sample regions and the surrounding areas, supporting sufficiently uniform infrared emissivity across the sample surfaces under the present measurement conditions. The images were acquired using a 1× infrared lens.

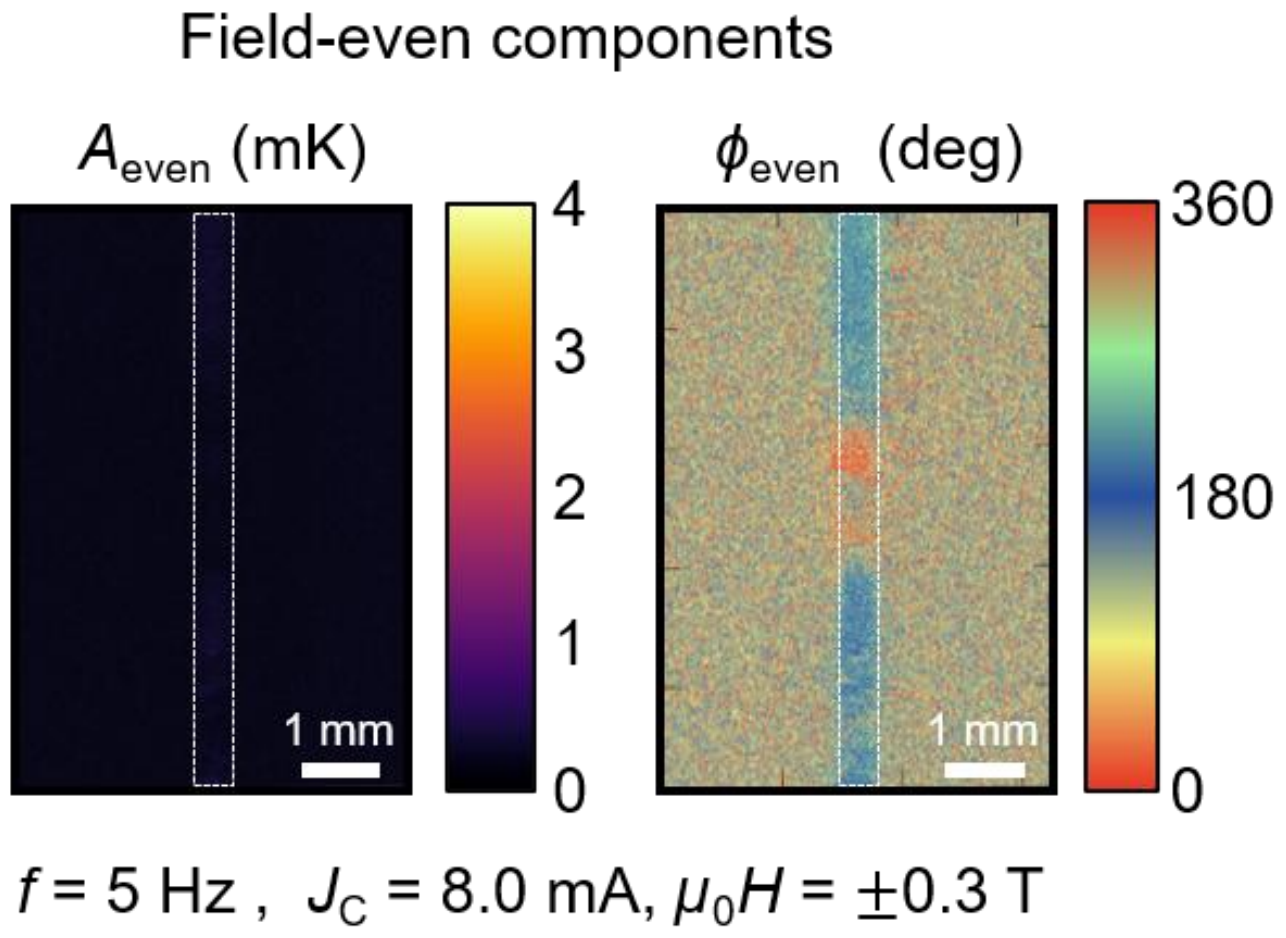

**Fig. S3. Field-even lock-in amplitude ($A_{even}$) and phase ($\phi_{even}$) images of the YIG/Pt (2.5 nm)/$CuO_x$ wedge (0–16 nm) sample.** Definitions are given in the MATERIALS AND METHODS. The same color scales as those in Fig. 2 of the main text were used. The white dashed boxes indicate the sample outline and are provided as guides to the eye. The $A_{even}$ values are comparable to the background fluctuation level and are typically less than 0.2 mK.

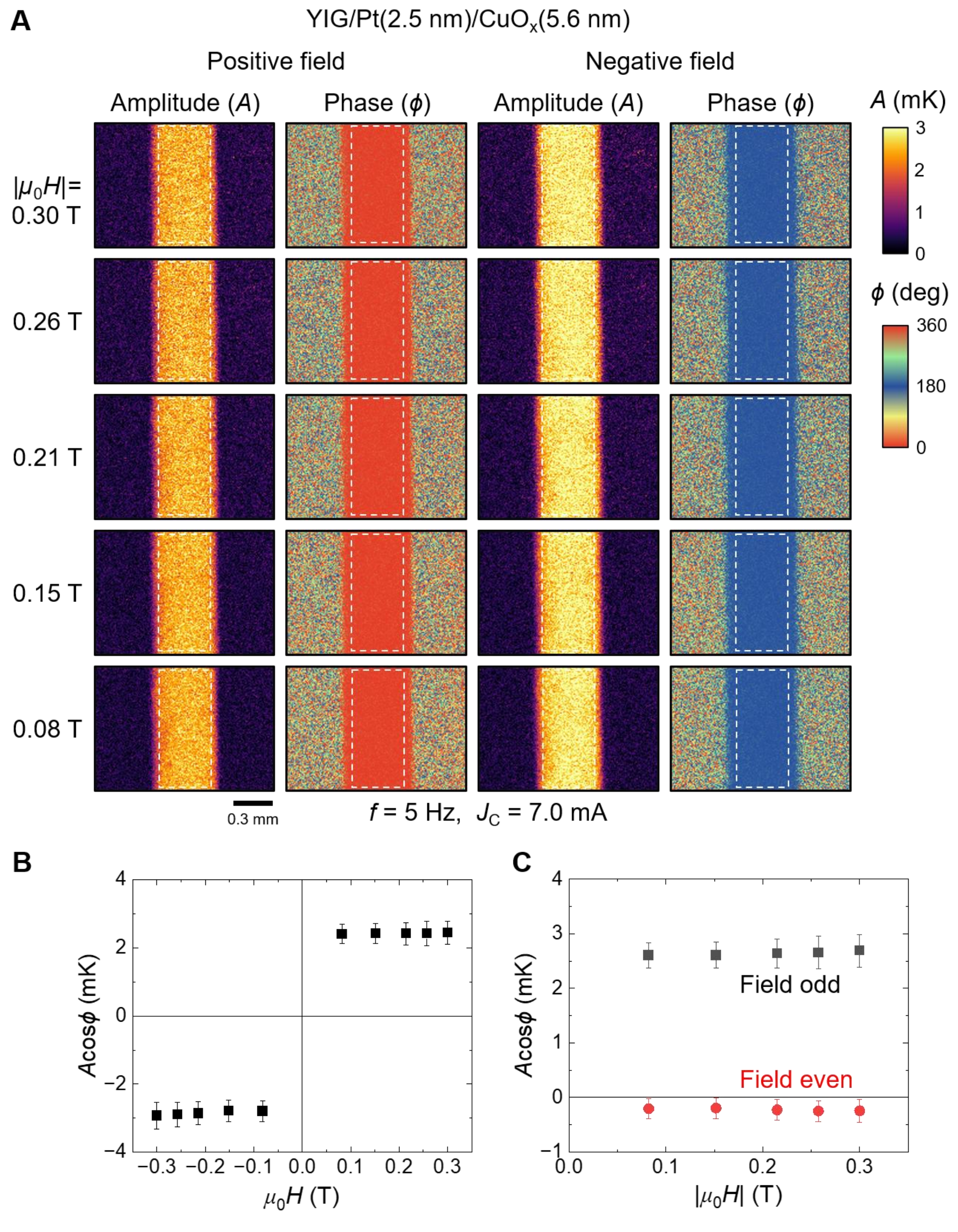

**Fig. S4. Magnetic-field dependence of the LIT signal in a YIG/Pt(2.5 nm)/$CuO_x$(5.6 nm) device.** (A) Raw lock-in amplitude ($A$) and phase ($\phi$) measured at each magnetic field ($\mu_0 H$). The measurements were performed at a fixed frequency of $f = 5$ Hz and an applied charge current of $J_C = 7.0$ mA. (B) Magnetic-field dependence of the in-phase temperature-modulation component, $A\cos\phi$. (C) Field-odd and field-even components obtained by antisymmetrization and symmetrization of the complex LIT signal. The LIT images were acquired using a 5× infrared lens.

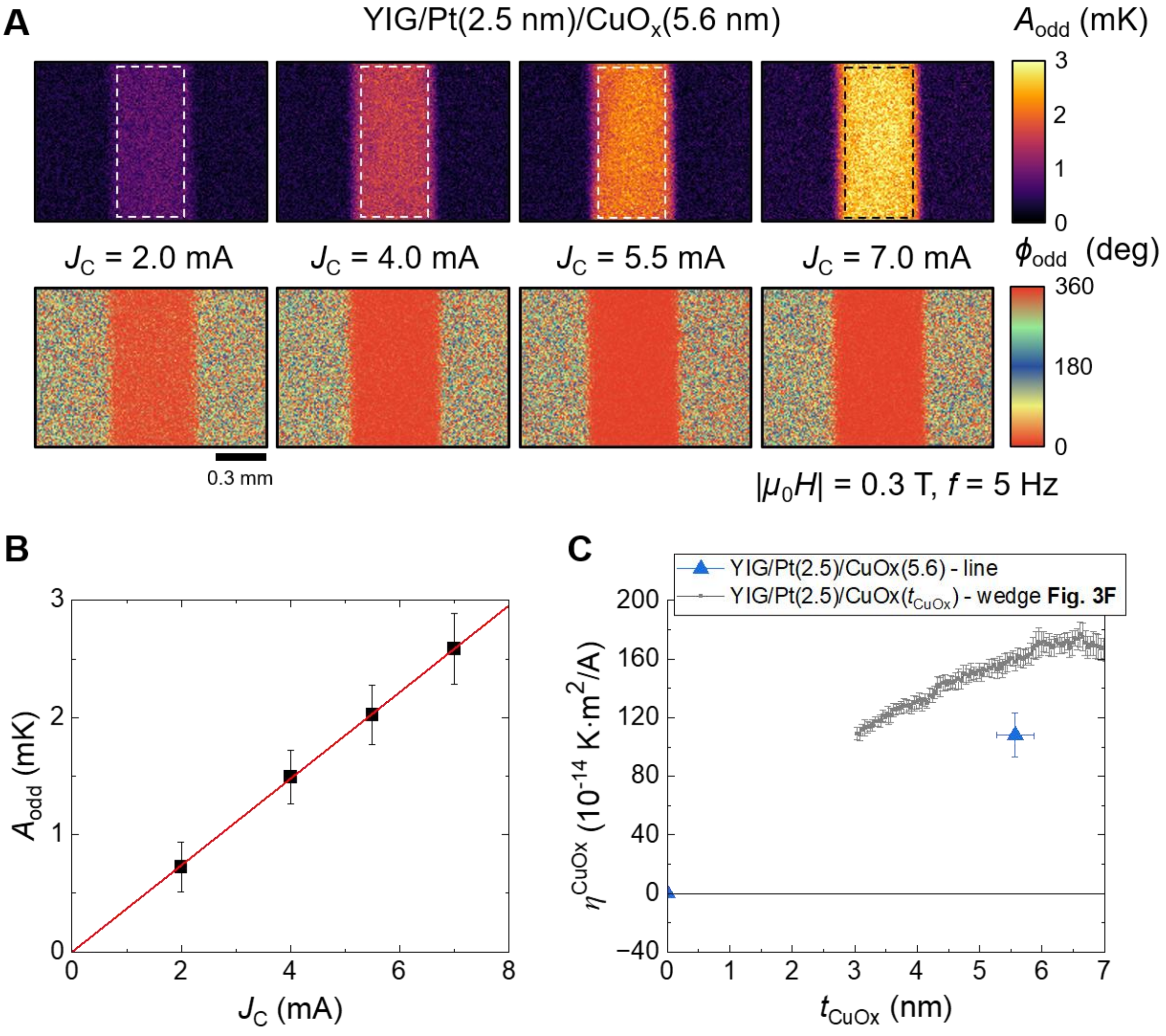

**Fig. S5. Applied-current dependence of the LIT signal in a YIG/Pt(2.5 nm)/$CuO_x$(5.6 nm) device.** (A) Field-odd lock-in amplitude ($A_{odd}$) and phase ($\phi_{odd}$) measured as a function of the applied charge current ($J_C$). The measurements were performed at a fixed frequency of $f = 5$ Hz and opposite magnetic fields $\mu_0 H = \pm 0.3$ T. (B) Applied-current dependence of $A_{odd}$. (C) Charge-to-heat conversion efficiency $\eta^{CuOx}$ associated with the $CuO_x$ orbital contribution, defined as the SPE-induced temperature modulation normalized by the current density flowing through the $CuO_x$ layer (see Equation 3 in the main text). The values for the wedge sample are replotted from Fig. 3F for comparison. The LIT images were acquired using a 5× infrared lens.

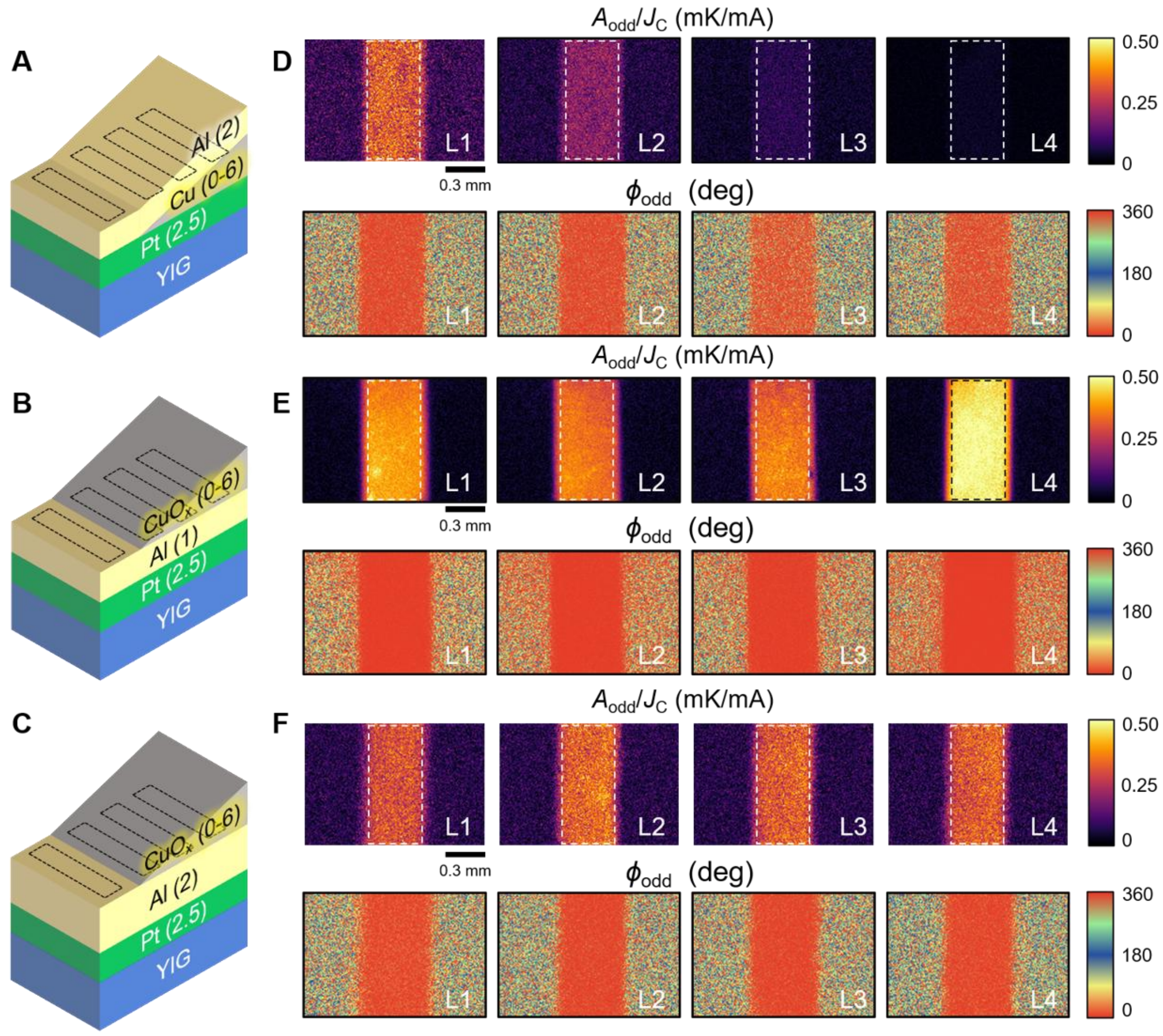

**Fig. S6. Schematic illustrations of the samples and measured lock-in thermography data.** (A)–(C) Schematic illustrations of the samples with (A) YIG/Pt(2.5 nm)/Cu(0–6 nm, wedge)/Al(2 nm), (B) YIG/Pt(2.5 nm)/Al(1 nm)/$CuO_x$(0–6 nm, wedge), and (C) YIG/Pt(2.5 nm)/Al(2 nm)/$CuO_x$(0–6 nm, wedge), corresponding to Fig. 4A–C of the main text, respectively. (D)–(F) Field-odd lock-in amplitude ($A_{odd}$) per applied current $J_C$ and phase ($\phi_{odd}$) for the samples. The values were obtained by averaging over the boxed regions. The LIT images were acquired using a 5× infrared lens, corresponding to a pixel size of 3 × 3 μm$^2$. Additional imaging and analysis details are provided in the MATERIALS AND METHODS.